\documentclass[12pt, a4paper]{article}

\usepackage{amsmath}
\usepackage{amsfonts}
\usepackage{amssymb}
\usepackage{graphicx, rotating}
\usepackage{epstopdf}
\usepackage{epsfig}
\usepackage{latexsym}
\usepackage{graphicx}
\usepackage{color}
\usepackage{amsmath,bm,amssymb}
\usepackage{cite}
\usepackage{slashed}
\usepackage{hyperref}
\hypersetup{colorlinks, citecolor=bluscuro, linkcolor=black, urlcolor=bluscuro}
\definecolor{rossos}{cmyk}{0,1,1,0.55}
\definecolor{bluscuro}{rgb}{0.15, 0.2, .85}
\definecolor{bluchiaro}{cmyk}{1,.3,0.,0.1}
\definecolor{brown}{rgb}{0.6, 0.14, 0.14}

\def\simlt{\stackrel{<}{{}_\sim}}
\def\simgt{\stackrel{>}{{}_\sim}}

\newcommand{\arXiv}[2]{\href{http://arxiv.org/pdf/#1}{{\tt [#2/#1]}}}
\newcommand{\arXivold}[1]{\href{http://arxiv.org/pdf/#1}{{\tt [#1]}}}

\newcommand{\be}{\begin{equation}}
\newcommand{\ee}{\end{equation}}
\newcommand{\bea}{\begin{eqnarray}}
\newcommand{\eea}{\end{eqnarray}}

\def\bma#1{\mbox{\boldmath{$#1$}}}

\begin{document}

\begin{titlepage}
\begin{flushright}
\end{flushright}
\vspace{.3in}

\vspace{1cm}
\begin{center}
{\Large\bf\color{black} The 750 GeV Diphoton Excess as a \\
\vspace*{0.2cm}
First Light on Supersymmetry Breaking}\\

\bigskip\color{black}
\vspace{1cm}{
{\large J.A. Casas$^a$, J.R.~Espinosa$^{b,c}$ and J.M. Moreno$^a$}
\vspace{0.3cm}
} \\[7mm]
{\it {$^a$ Instituto de F\'{\i}sica Te\'orica, IFT-UAM/CSIC, Nicol\'as Cabrera 13, UAM Cantoblanco, 28049 Madrid, Spain. }}\\
{\it {$^b$Institut de F\'isica d'Altes Energies (IFAE), The Barcelona Institute of Science and Technology (BIST), Campus UAB, E-08193, Bellaterra (Barcelona), Spain}}\\
{\it $^c$ {ICREA, Instituci\'o Catalana de Recerca i Estudis Avan\c{c}ats, Barcelona, Spain}}
\end{center}
\bigskip

\vspace{.4cm}

\begin{abstract}

One of the most exciting explanations advanced for the recent diphoton excess found by ATLAS and CMS is in terms of sgoldstino decays: a signal of low-energy supersymmetry-breaking scenarios.  The sgoldstino, a scalar, couples directly to gluons and photons, with strength related to gaugino masses, that can be of the right magnitude to explain the excess. However, fitting the suggested resonance width, $\Gamma\simeq 45$ GeV, is not so easy. In this paper we explore efficient possibilities to enhance the sgoldstino width, via the decay into two Higgses, two Higgsinos and through mixing between the sgoldstino and the Higgs boson. In addition, we present an alternative and more efficient mechanism to generate a mass splitting between the scalar and pseudoscalar components of the sgoldstino, which has been suggested as an interesting alternative explanation to the apparent width of the resonance.

\end{abstract}
\bigskip

\end{titlepage}

 
\section{Introduction\label{sec:intro}} 

The ATLAS and CMS collaborations have recently reported an excess in diphoton searches at $\sqrt{s}=13$ TeV for a $\sim$ 750 GeV diphoton invariant mass\cite{data, Aad:2015mna,CMSgg8}. The local significance is $3.9\sigma$ (ATLAS) and $2.6\sigma$ (CMS), although it gets smaller once the  look-elsewhere effect is taken into account. However, the fact than both experiments see the signal in the same place has created in the community the expectation that it could be the long expected signal of new physics. 

Once the accumulated statistics at ATLAS and CMS grow large enough, we will see finally whether or not  this excess is an statistical fluctuation. In the meantime, it  is tempting to try and interpret the data as a signal of new physics as the flood of papers studying different BSM scenarios that could accommodate the excess testify. Those most relevant to our discussion are \cite{Strumia,Javi,Torre,ovs,Ellis,photonfusion,Adam}.
In our opinion, probably the most exciting theoretical possibility to accommodate this resonance is the one pursued by the authors of \cite{Javi,Torre,ovs}, who have contemplated scenarios with a scale of SUSY breaking not far from the TeV scale (low-scale SUSY breaking) \cite{BCEN,CEH,Seiberg,otherstudies}.  Potentially, these models contain the main ingredient to fit the signal: an scalar field $\phi$ (the sgoldstino) coupled to gluons and photons in a direct way, so that an effective production via gluon fusion and the subsequent decay into photons are possible. Beside reproducing the observed cross section, any good explanation of the data should account for the apparent sizeable width of the resonance, $\Gamma_\phi/M_\phi\simeq 0.06$, although the data are not yet conclusive. The authors of  ref. \cite{Torre} discussed a simple explanation for the apparent width: a mass splitting (as advocated in \cite{Strumia}) between the scalar and pseudoscalar degrees of freedom of the sgoldstino.

 In this paper we review the explanation of the diphoton signal (sect.~\ref{sec:excess}) based on this type of scenarios (sect.~\ref{sec:LSSB}), exploring mechanisms for a broad  $\Gamma_\phi$, potentially consistent with the data.  We present other mechanisms for the mentioned sgoldstino mass splitting, which are more efficient than those considered up to now  (sect.~\ref{sec:two}). In our analysis we  discuss some subtleties not previously considered that can constrain and affect substantially the results. We also discuss the possibility that sgoldstinos decay efficiently into Higgses (sect.~\ref{sec:hh}), as the partial width into that channel is naively parametrically enhanced with respect to other channels; into Higgs decay channels through sgoldstino-Higgs mixing (sect.~\ref{sec:shmix}); and into Higgsinos (sect.~\ref{sec:other}), as there is more freedom to enhance this width without clashing with previous LHC searches.

\section{The low-scale SUSY-breaking scenario\label{sec:LSSB}}

The low-scale SUSY-breaking (LSSB)  scenario \cite{BCEN,CEH,Seiberg,otherstudies} is a framework in which
the scale of SUSY breaking, $\sqrt{F}$, and its mediation, $M$, are not far from the TeV scale. The main differences with respect to more conventional supersymmetric models, where the latter scales are large, are the following:  i) the low-energy effective theory includes the chiral superfield, $\Phi$, responsible for SUSY breaking, in particular its fermionic (goldstino) and scalar (sgoldstino) degrees of freedom; ii) besides the ordinary SUSY-soft breaking terms, the effective theory contains additional hard-breaking operators, e.g. quartic Higgs couplings. The latter make the Higgs sector resemble a two-Higgs doublet model with an additional (complex) singlet. LSSB models present a much milder electroweak fine-tuning than usual MSSMs \cite{CEH, Seiberg} and a rich phenomenology \cite{BCEN,CEH,Seiberg,otherstudies}.
As discussed in refs.\cite{Javi,Torre,ovs}, the LSSB scenario can nicely explain the diphoton excess at 750 GeV observed at the LHC. 

Let us summarize the main ingredients of LSSB scenarios. Expanding in inverse powers of $M$, superpotential, $W$,  K\"ahler potential, $K$, and  the gauge kinetic function, $f_{ab}$, read \cite{BCEN}
\bea
\label{W}
W&= &W_{\rm MSSM} + F\left(\Phi+ \frac{\rho_\phi}{6M^2} \Phi^3+\cdots\right) + \left(\mu + \frac{\mu'}{M}\Phi+\cdots\right) H_u\cdot H_d
\nonumber\\
&+& \frac{1}{2M}\left(\ell+\frac{\ell'}{M}\Phi+\cdots\right) (H_u\cdot H_d)^2 \ +\ \cdots\ ,\\
\label{K}
K&=&|\Phi|^2\left(1-\frac{\alpha_\phi}{4M^2}|\Phi|^2+\cdots\right) + |H_u|^2\left[1+\frac{\alpha_u}{M^2}|\Phi|^2+\cdots\right]+ |H_d|^2\left[1+\frac{\alpha_d}{M^2}|\Phi|^2+\cdots\right]
\nonumber\\
&+& \left[H_u\cdot H_d\left(\frac{\alpha_{ud}}{2M^2}\bar\Phi^2+\cdots\right) +\ {\rm h.c.}\right]\ +\ \cdots\ ,\\
%
\label{fab}
f_{ab}&=&\frac{\delta_{ab}}{g_a^2}\left[1+c_a\frac{\Phi}{M}\ +\ \cdots\right] \ .
\eea
%
%
Here all the parameters are dimensionless, except the $\mu,\mu'\cdots$ parameters in the superpotential, which have dimensions of mass.
Replacing $\Phi$ by its auxiliary field, $F$, one gets the soft breaking terms of the theory.
In particular, from Eq.~(\ref{fab}), one gets masses for gluinos, $M_3$, winos, $M_2$, and the bino, $M_1$, e.g. $M_1=c_1F/M$. 
Likewise, replacing $\Phi$ by its scalar component, a complex singlet field, that we also denote by  $\Phi$,
\be
\Phi=\frac{1}{\sqrt{2}}(\phi_S + i \phi_P)\ ,
\label{twophis}
\ee
(where $\phi_S$ is the scalar component and  $\phi_P$ the pseudoscalar one), one obtains couplings of the $\phi$'s with the MSSM fields. In particular, the coupling to gluons and photons is directly related to gaugino masses as:
\bea
\label{XGG}
{\cal L}\supset \frac{M_3}{2\sqrt{2} F} \ {\rm tr}\  G_{\mu\nu}^a(\phi_S G^{a\mu\nu} -\phi_P \tilde G^{a\mu\nu}) \ +\ \frac{M_{\tilde \gamma}}{2\sqrt{2} F} \ {\rm tr}\ F_{\mu\nu}(\phi_S F^{\mu\nu}-\phi_P\tilde F^{\mu\nu})\ ,
\eea
where $M_{\tilde \gamma}$ is the photino mass, 
\bea
\label{Mgamma}
M_{\tilde \gamma}=M_1 \cos^2 \theta_W +M_2 \sin^2 \theta_W \ .
\eea

Similarly,  from Eqs.~(\ref{W}) and  (\ref{K}), the scalar potential $V=V_F + V_D$ for the two supersymmetric Higgs doublets plus the complex singlet field $\Phi$, is:\footnote{A linear term in $\Phi$ can always be removed by a field redefinition.  For more details, see \cite{BCEN}.}
\bea
\label{V}
V & = & F^2 + \alpha_\phi \tilde{m}^2 |\Phi|^2 
+  {1\over 2}(\rho_\phi  \tilde{m}^2 \Phi^2  + {\rm h.c.})
+ m_{H_u}^2 |H_u|^2 +  m_{H_d}^2 |H_d|^2 +  \left(m_{12}^2
H_u\cdot H_d + {\rm h.c.}\right) \nonumber\\ & + & (m_{X_1} \Phi +
m_{X_1}^*\Phi^* )|H_u|^2  + (m_{X_2} \Phi + m_{X_2}^*\Phi^* )
 |H_d|^2 +
\left[ (m_{X_3} \Phi + m_{X_4} \Phi^* ) H_u\cdot H_d + {\rm h.c.} \right]
\nonumber\\ & + & {1\over 2} \lambda_1 |H_u|^4 + {1\over 2} \lambda_2
|H_d|^4  +  \lambda_3 |H_u|^2 |H_d|^2 + \lambda_4 |H_u\cdot H_d|^2
\nonumber\\ & + & \left[ {1 \over 2} \lambda_5 \left(H_u\cdot
H_d\right)^2   + \lambda_6 |H_u|^2 H_u\cdot H_d  + \lambda_7 |H_d|^2
H_u\cdot H_d + {\rm h.c.} \right]+\ldots \eea
where the dots stand for  higher order terms in $\Phi$ and  nonrenormalizable terms suppressed by powers of $M$. The  various mass parameters and quartic couplings in (\ref{V}) are explicit combinations of the parameters in $W$ and $K$ (see ref.\cite{BCEN} for explicit formulae).
%
%
As a summary, denoting by $\mu$ the typical scale of the supersymmetric mass parameters [$\mu, \mu', \cdots$ in Eq.~(\ref{W})] and $\tilde m=F/M$, the mass terms in the potential have contributions of order $\mu^2$, $ \tilde{m}^2$, $\tilde{m}\mu$. We assume $\mu \simlt\tilde m$, so that all these squared mass terms are expected to be $\simlt\tilde m^2$. Analogously, trilinear terms, $m_{X_i}$, have contributions of order $\mu^2/M$, $ \tilde{m}^2/M$, $\tilde{m}\mu/M$. Finally, the Higgs quartic couplings have supersymmetric $D$-term and $F$-term contributions, where the latter include supersymmetry breaking contributions as well: 
$\lambda_i = \lambda_i^{(D)} +  \lambda_i^{(F)}$. The
$\lambda_i^{(D)}$ are as in the MSSM: 
\be
\lambda_1^{(D)}=\lambda_2^{(D)}={1\over 4}(g^2 + {g'}^2) \ , \;\;
\lambda_3^{(D)}={1\over 4}(g^2 - {g'}^2) \ , \;\;
\lambda_4^{(D)}=-{1\over 2}g^2 \ ,
\label{H4D}
\ee 
and $\lambda_5^{(D)}=\lambda_6^{(D)}=\lambda_7^{(D)}=0$. Besides, typically $ \lambda_i^{(F)}\sim \tilde{m}^2/M^2,\tilde{m}\mu/M^2,\mu^2/M^2$, although some of these couplings can receive contributions at a lower order, $\lambda_{5}^{(F)}\sim \tilde m/M$,
$\lambda_{i=6,7}^{(F)}\sim \mu/M$.  In what follows we will generically assume $ \lambda_i^{(F)}\sim \tilde{m}^2/M^2$ but the reader shuld keep in mind this exception, which might be important in some cases.
The effective quartic  self-coupling of the light (SM-like) Higgs, $\lambda|H|^4$, reads
\be
\lambda = \lambda^{(D)} + \lambda^{(F)} +\delta_{\rm rad}\lambda\ ,
\ee
where
\be
\lambda=\frac{1}{2}\left(\lambda_1 c_\beta^2+ \lambda_2s_\beta^2\right) + \frac{1}{4}\left(\lambda_3+\lambda_4+\lambda_5\right) \sin^2 2\beta +\left(\lambda_6 c_\beta^2+ \lambda_7s_\beta^2\right)\sin2\beta\ ,
\ee
with $\tan\beta=\langle H_u\rangle/\langle H_d\rangle\equiv v_u/v_d$. This quartic coupling determines the Higgs mass as in the SM, i.e. $m_h^2=2\lambda v^2$, with $v^2=v_u^2+v_d^2=(246\ {\rm GeV})^2$. The sizes of the various contributions are
\be
2\lambda^{(D)}v^2=m_Z^2\cos^2(2\beta)\ ,\quad
2\lambda^{(F)}v^2\sim \frac{\tilde{m}^2}{M^2}v^2\ ,\quad
2\delta_{\rm rad}\lambda v^2 \sim  \frac{3}{2\pi^2}\frac{m_t^4}{v^2}
\log\frac{m_{\tilde{t}}^2}{m_t^2}
\ ,
\ee
where $m_{\tilde{t}}$ is the stop mass scale.

The $\lambda^{(F)}$ contributions to the Higgs quartic coupling 
play a crucial role in this kind of scenario: moderate values of the 
ratio $\tilde{m}/M$ or $\tilde{m}^2/M^2$ ($\sim 0.1-0.2$ for large $\tan\beta$) can push up the Higgs mass significantly, so that the measured value $m_h\simeq 125$ GeV can be achieved alleviating the naturalness tension of supersymmetric models \cite{CEH}. 

In this paper we are interested in the decoupling regime with a single light Higgs doublet. The singlet $\Phi$ should have a 750 GeV mass while a linear combination of $H_1$ and $H_2$ will have mass around the TeV scale.

\section{The diphoton excess\label{sec:excess}}

\begin{figure}[t]
$$\includegraphics[width=0.6\textwidth]{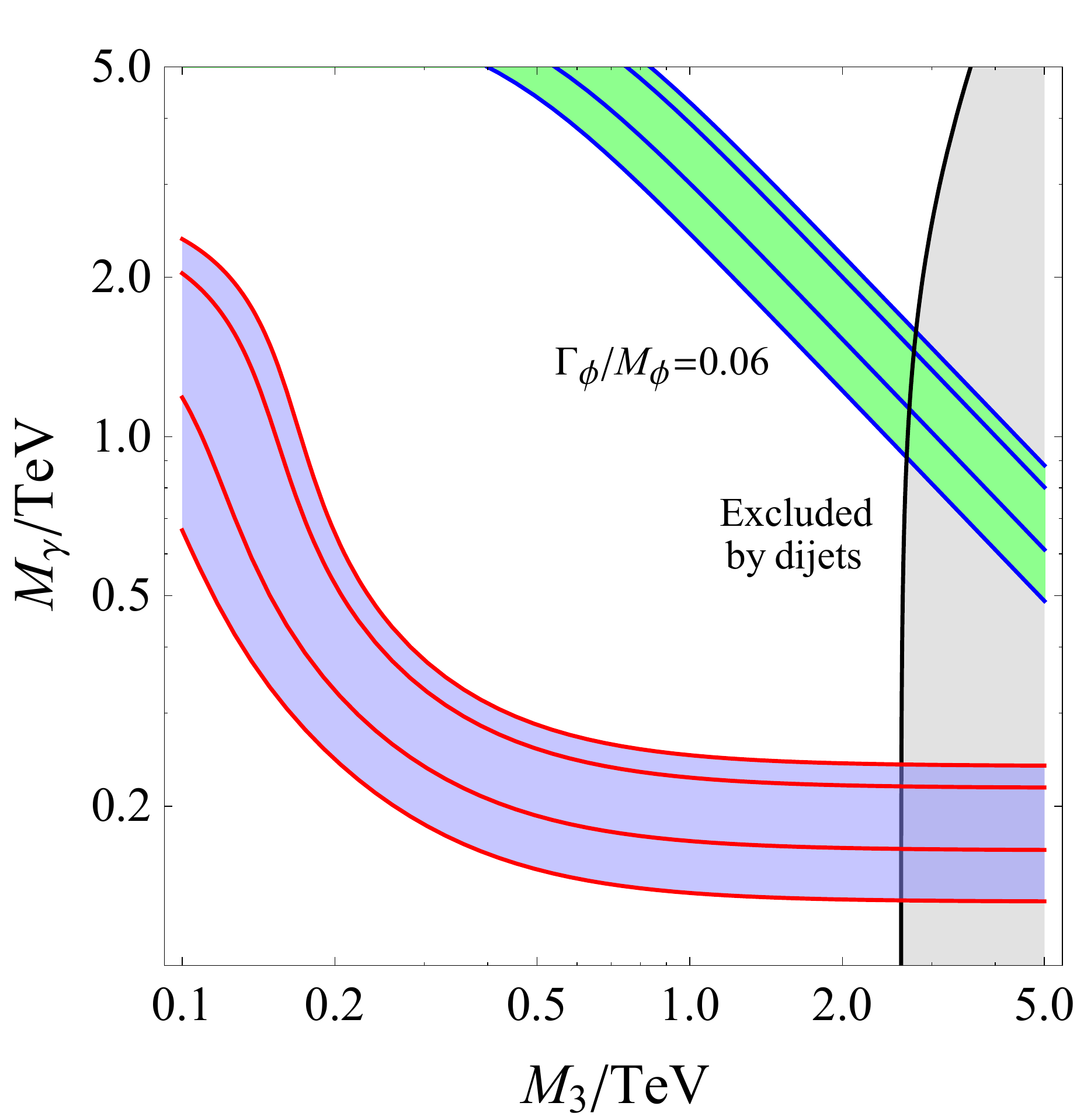} $$     
\begin{center}
\caption{Regions of the ${M_3,M_{\tilde \gamma}}$ plane in which the observed cross section $\sigma(pp\rightarrow \phi \rightarrow \gamma\gamma)$ is reproduced, assuming $\sqrt{F}=4$ TeV and $M_1=M_2$. The green and blue bands correspond to $\Gamma=0.06 M_\phi$ and the  actual $\Gamma$ from decays into SM gauge bosons, respectively. Thin (broad) bands correspond to $1\sigma$ ($2\sigma$). The gray region is excluded by LHC dijet searches (but applies only to the blue band).
\label{fig:M3Mph}}
 \end{center}
\end{figure}

The total cross section, $\sigma(pp\rightarrow \phi \rightarrow \gamma\gamma)$, where $\phi$ generically denotes either $\phi_S$ or $\phi_P$, is dominated by gluon fusion, and can be expressed in terms of the partial widths, $\Gamma_{gg}, \Gamma_{\gamma \gamma}$ and total width $\Gamma$, as
\be
\sigma(pp\rightarrow \phi \rightarrow \gamma\gamma) \simeq \frac{1}{sM_\phi\Gamma} (C_{gg}\Gamma_{gg}+C_{\gamma\gamma}\Gamma_{\gamma\gamma}) \Gamma_{\gamma\gamma},
\ee
with $C_{gg}\simeq 2137$ , $C_{\gamma\gamma}\simeq 54$ at 13 TeV (see e.g. \cite{Strumia}). We include in the production mechanism photon-fusion, which can also be relevant, as discussed in \cite{photonfusion}.
 From Eq.~(\ref{XGG}), one can extract the partial widths of $\phi$ into gluons and photons:
\be
\label{Gammas}
\Gamma_{ gg}=\frac{M_3^2M_\phi^3}{4\pi F^2},\ \ \ 
\Gamma_{ \gamma\gamma}=\frac{M_{\tilde\gamma}^2M_\phi^3}{32\pi F^2}\ .
\ee
Fig.~\ref{fig:M3Mph} shows the regions in the $\{M_3,M_{\tilde \gamma}\}$ plane where $\sigma(pp\rightarrow \phi \rightarrow \gamma\gamma)$  (summing the contributions from both $\phi_S$ and $\phi_P$) is consistent with the combined experimental value, which we naively estimate as  $\sigma\sim 8\pm2.1$ fb, for a typical value $\sqrt{F}=4\ {\rm TeV}$.\footnote{If we impose $M_i<\sqrt{F}$ as an absolute limit required for the EFT expansion to make sense, explaining the observed diphoton cross-section implies $\sqrt{F}\simlt 8$ TeV. The allowed values of $M_3, M_{\tilde\gamma}$ simply scale as $M\propto F$. } The green band corresponds to the assumption (favored by ATLAS) that the total width is $\Gamma=0.06\ M_\phi$. The blue band corresponds to the total width calculated summing $\Gamma_{gg}+\Gamma_{\gamma\gamma}$ plus other contributions  (see e.g. \cite{Torre2}) from the decay into $WW$, $ZZ$, $Z\gamma$
(choosing $M_1=M_2$), which can be comparable to $\Gamma_{\gamma\gamma}$, while $\phi$-decays into tops or goldstinos give only a small contribution.
The region consistent with the data corresponds to the area between these two bands. Finally the gray region is excluded by LHC dijet searches \cite{dijets} (with the boundary value of $M_3$ scaling as $F$).  This exclusion limit has been calculated assuming that the $\phi$ width is determined by $\phi$ decays to SM gauge bosons only and therefore applies only to the blue band.

Playing only with the decay channels discussed  above one cannot explain $\Gamma=0.06 M_\phi$, as is clear from the figure: the band intersection is excluded by dijet searches.
Therefore, in order to get closer to the apparently favored green band, there are two possibilities (apart from the possibility that  the evidence for the broad width eventually dissapears, even if the resonance is there): that the experimental data correspond in fact to two unresolved resonances, mimicking a broad width; or that other decays of $\phi$ enhance its width by the right amount. We explore these possibilities in the following sections.

\section{Two unresolved resonances\label{sec:two}}

Although the mass resolution in the diphoton channel is very good, with the current statistics, two narrow resonances close in mass could well be responsible for the apparently wide resonance that ATLAS reports. In the sgoldstino scenario we consider, such double resonance is a natural possibility, as the complex singlet field $\Phi$ has two real components, as explicitly shown in Eq.~(\ref{twophis}), and generic scalar potentials give different masses to $\phi_S$ and   $\phi_P$. Indeed, such mass splitting has been proposed in \cite{Strumia,Torre} as a resolution to the puzzle of the large width of the 750 GeV resonance. In this section we go beyond that analysis in 
several respects, pointing out that other sources of sgoldstino mass splitting, different from the one considered in \cite{Torre}, are possible and interesting.

After electroweak symmetry breaking (EWSB), the mass matrix for the neutral scalars generically mixes the two sgoldstino fields $\phi_{S,P}$ with three Higgs fields: the light Higgs $h^0$, and the two heavy ones, $H^0$ and $A^0$. In first approximation, neglecting effects from EWSB, as $v\ll M_\phi$,  one simply gets from (\ref{V})
the two sgoldstino squared-mass eigenvalues $\tilde{m}^2(\alpha_\phi\pm \rho_\phi)$. A small mass splitting requires $\rho_\phi\ll \alpha_\phi$, in which case $\Delta M_\phi \simeq \tilde{m} \rho_\phi/\sqrt{\alpha_\phi}\simeq M_\phi \rho_\phi/\alpha_\phi$. So, $\Delta M_\phi\sim 30$ GeV requires the mild hierarchy $\rho_\phi/\alpha_\phi\sim 0.04$ between
$\rho_\phi$ and $\alpha_\phi$, the Wilson coefficients of the first nonrenormalizable terms of $\Phi$ in the K\"ahler potential and superpotential, see Eqs.~(\ref{W}) and (\ref{K}). This source of sgoldstino mass splitting, not considered in  \cite{Torre}, is potentially the largest one.

Additional, or alternative, sources of sgoldstino mass splitting from EWSB effects can have two different origins:  {\it (i)}  the trilinear couplings that connect the singlet $\Phi$ and the Higgs doublets $H_i$  in Eq.~(\ref{V}), and {\it (ii)} mixed quartic couplings, of the type $(\lambda_a\Phi^2+\lambda_a^*\Phi^{*2})(a_i|H_i|^2+b\ H_u\cdot H_d+h.c.)]$, that we did not write explicitly in (\ref{V}). We consider them in turn.

Type {\it (i)} splitting occurs through contributions to the off-diagonal entries, $h_i-\phi_{S,P}$ (with $h_i\equiv h^0,H^0,A^0$), in the Higgs-sgoldstino mass matrix, that are different for 
$\phi_S$ and $\phi_P$ and are of order $m_X v$, with $m_X\sim \tilde{m}^2/M$.\footnote{The trilinear couplings also induce a small vacuum expectation value $\langle \phi\rangle\sim m_X v^2/M_\phi^2$ that plays a subdominant role in the discussion that follows.}   
The size of the $\phi$ mass splitting depends on the kind of Higgs that mixes with the sgoldstinos. 

First, if sgoldstinos mix with the light Higgs via a $\delta V=\frac{1}{2}m_X\phi h^2$ term in the potential,
there are two potentially dangerous side-effects. The mixing leads, via eigenvalue repulsion, to a reduction of the light Higgs mass by $\delta m_h^2\sim m_X^2 v^2/M_\phi^2$ (which should be bounded to be naturally smaller than $\kappa m_h^2$, with $\kappa$ of order a few), and  an upward shift of $M^2_{\phi_S}$ and $M^2_{\phi_P}$  of the same order $\delta m_h^2$ but not necessarily equal for $\phi_S$ and $\phi_P$, resulting in a $\Delta M_\phi^2$  that is a fraction of  $\delta m_h^2$.
Noting that $m_X\sim \tilde{m}^2/M$, and $M_\phi\sim \tilde{m}$, one gets $\Delta M_\phi\sim \tilde{m}(v^2/M^2)$. This is the kind of mass splitting discussed in \cite{Torre}. Using the above-mentioned natural constraint $\delta m_h^2\simlt \kappa m_h^2$, we get  $\Delta M_\phi\simlt m_h^2/(2M_\phi)\sim 10\kappa$ GeV. 
%
Second, since the light Higgs picks up a small sgoldstino component, this mixing reduces universally the Higgs couplings  (up to the small couplings of the sgoldstino to different SM particle species). This reduction of Higgs couplings, which has an effect similar to an invisible Higgs width, is bounded by LHC Higgs data to $\sin^2\alpha\simlt 0.2$ at 95\% C.L. \cite{FRU}, where $\alpha$ is the sgoldstino-Higgs mixing angle, given by
\be
\sin2\alpha=\frac{2m_X v}{M_\phi^2-m_h^2}\ .
\label{alpha}
\ee
This bound roughly translates into a bound on the splitting, $\Delta M_\phi\simlt (M_\phi/2)\sin^2 \alpha$. In general, splittings $\Delta M_{\phi}\sim{\rm few}\times 10$ GeV imply $\sin^2\alpha \simgt 0.1$, which could be visible in the future.  However, as we show in section~\ref{sec:shmix}, the mixing angle is strongly constrained by other physical effects, which casts doubts on the viability of this option.
%

If the sgoldstinos mix instead with the heavy Higgs doublet, of mass $M_H$, the sgoldstino mass 
splitting depends on the relative size of $M_\phi$ and $M_H$.   If $M_H\gg M_\phi$, one gets $\Delta M_\phi^2\sim m_X^2v^2/M_H^2$, smaller than the splitting in the previous case. The case $M_H\ll M_\phi$ would lead to mass splittings similar to those already considered but is difficult to realize due to constrains from heavy Higgs searches. Finally, if $M_H\simeq M_\phi$, one gets instead $\Delta M_\phi^2\sim m_X v$, and then $\Delta M_\phi\sim m_X v/M_\phi\sim \tilde{m}(v/M)$, parametrically larger than previous splittings. Notice that in this latter case the mixing between the sgoldstino(s) and the heavy Higgs doublets can be significant, with potentially important implications for the sgoldstino decays: the total width of the sgoldstino would be affected by the large fermionic width of the heavy Higgses, which can be of order $\sim 10$ GeV in that mass range.

In the type {\it (ii)} case, there are  EWSB contributions to the $\phi_S-\phi_P$ entries of the mass matrix, of order $\lambda_X v^2$,
with $\lambda_X\sim \tilde{m}^2/M^2$. The sgoldstino mass splitting results either from contributions to off-diagonal squared-mass entries or from different contributions to the diagonal entries. More precisely,
$(\lambda_a\Phi^2+\lambda_a^*\Phi^{*2})(a_i|H_i|^2+b\ H_u\cdot H_d+h.c.)$ leads to $\Delta M_\phi^2\simeq 2|\lambda_a|(a_i v_i^2+2bv_uv_d)$, with ${\rm Re}\lambda_a$ (${\rm Im}\lambda_a$) contributing to the (off-)diagonal splitting.
The generic result is $\Delta M_\phi^2\sim \lambda_X v^2$ and somewhat sizeable values of $\lambda_X$ are required for an sgolstino mass splitting of the right size: {\it e.g.} $\lambda_X\sim 0.5-0.75$  for $\Delta M_\phi\sim 20-30$ GeV.

\section{A large $\bma\phi$-width via $\bma{hh}$ decay?\label{sec:hh}} 

In the previous section we have discussed several ways to generate naturally an sgoldstino mass splitting that can explain the large width of the 750 GeV diphoton resonance. In particular, we presented several options beyond the one discussed in \cite{Torre}, which was based on a trilinear coupling between the sgoldstinos and the light Higgs. In fact, such trilinear couplings open a new decay channel for the sgoldstinos into two light Higgses, $\phi\rightarrow hh$, with a partial width that is parametrically large and can play a central role in determining the total $\phi$-width.

\begin{figure}[t]
$$\includegraphics[width=0.4\textwidth]{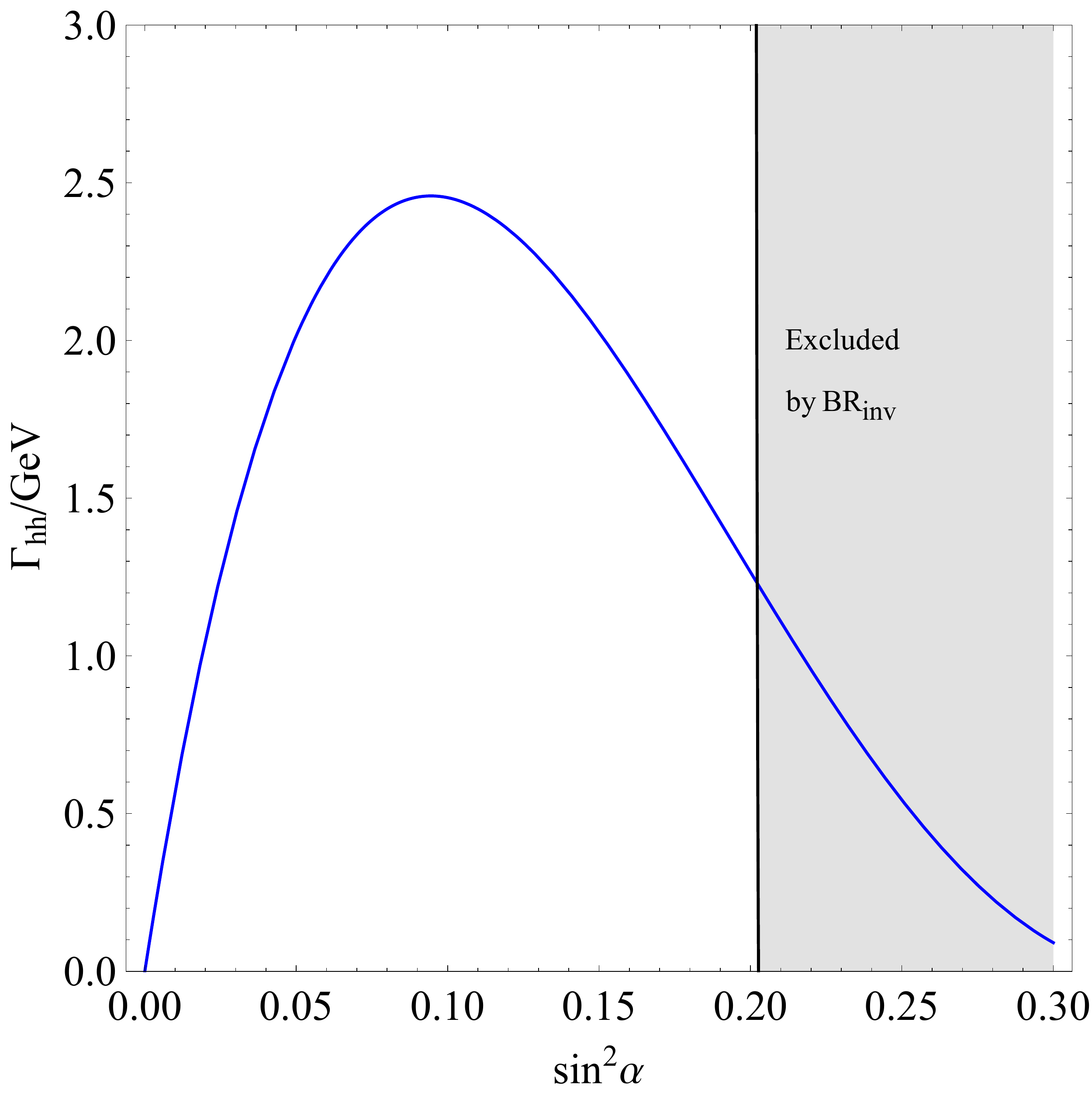} $$     
\begin{center}
\caption{Partial width of sgoldstino decay into two light Higgses (induced by a trilinear coupling $m_X\phi h^2/2$) as a function of the sgoldstino-Higgs mixing angle $\alpha$. The gray region is excluded by the LHC limit on the Higgs invisible width.
\label{fig:hh}}
 \end{center}
\end{figure}

Let us write schematically the relevant trilinear couplings as 
\be
\delta V=\frac{1}{2}m_X\phi h^2\ ,
\label{trilinear}
\ee
where we generically denote the (real) sgoldstinos as $\phi$ and $m_X\sim \tilde{m}^2/M$ as usual.  Through this coupling,
$\phi$  can decay to two light Higgs 
bosons\footnote{In addition to (\ref{trilinear}), nonrenormalizable  operators that one expects to appear in the effective theory \cite{BCEN},  like $\delta{\cal L}_5 = \kappa_i\partial^\mu \Phi \ (H_i^\dagger \overleftrightarrow{D_\mu} H_i)/M +\kappa_i' \partial^\mu \Phi \ \partial_\mu |H_i|^2/M+ h.c.$,
could potentially contribute to the decay $\phi\rightarrow hh$. However, the first operator in $\delta{\cal L}_5$ does not contribute to the decay of $\Phi$ into two on-shell Higgses and the second one can be rewritten, by integration by parts and use of the equations of motion of $\Phi$, as an operator of the same form as (\ref{trilinear}) with a coefficient $\sim M_\phi^2/M$, which is of the same order as $m_X$. 
}. 
Naively, using a large enough $m_X$, one can get a sizeable partial width ({\em e.g.} $m_X\sim 1.9$ TeV to get $\Gamma_\phi/M_\phi=0.06$). However, there is an obstruction to how large $m_X$ can be: as  we saw in the previous section, this coupling also  induces an sgoldstino-Higgs mixing with angle $\alpha$ given by Eq.~(\ref{alpha}), $\sin2\alpha=2m_Xv/(M_\phi^2-m_h^2)$. This imposes the upper bound $m_X\leq (M_\phi^2-m_h^2)/(2v)\simeq 1.1$ TeV, which becomes $m_X\simlt 0.9$ TeV once the LHC upper bound $\sin^2\alpha\simlt 0.2$ is imposed.  This will limit how large the $\Gamma_{hh}$ can be. In addition, in order to calculate $\Gamma_{hh}$, we have to re-write Eq.~(\ref{trilinear}) in terms of mass eigenstates, and this introduces mixing-angle factors. Then the coupling relevant for the sgolsdtino decay is not $m_X$ but $m_X (c^3_\alpha - 2 c_\alpha s^2_\alpha)$, with $s_\alpha=\sin\alpha$, etc., so  the partial width reads
\be
\label{Gammahh}
\Gamma_{hh} = \frac{1}{32\pi} (c^3_\alpha - 2 c_\alpha s^2_\alpha)^2\frac{m_X^2}{M_\phi}\sqrt{1-4m_h^2/M_\phi^2} \ .
\ee
As $m_X$ is related to the mixing angle by Eq.~(\ref{alpha}), the partial width is uniquely determined by $\alpha$. This is shown in Fig.~\ref{fig:hh}, which makes clear that the naive expectations are not fullfilled, and the maximal partial width is $\sim 2.5$ GeV (corresponding to $m_X\sim 700$ GeV), too small to explain the large value of $\Gamma_\phi$ suggested by ATLAS.

\begin{figure}[t]
$$\includegraphics[width=0.6\textwidth]{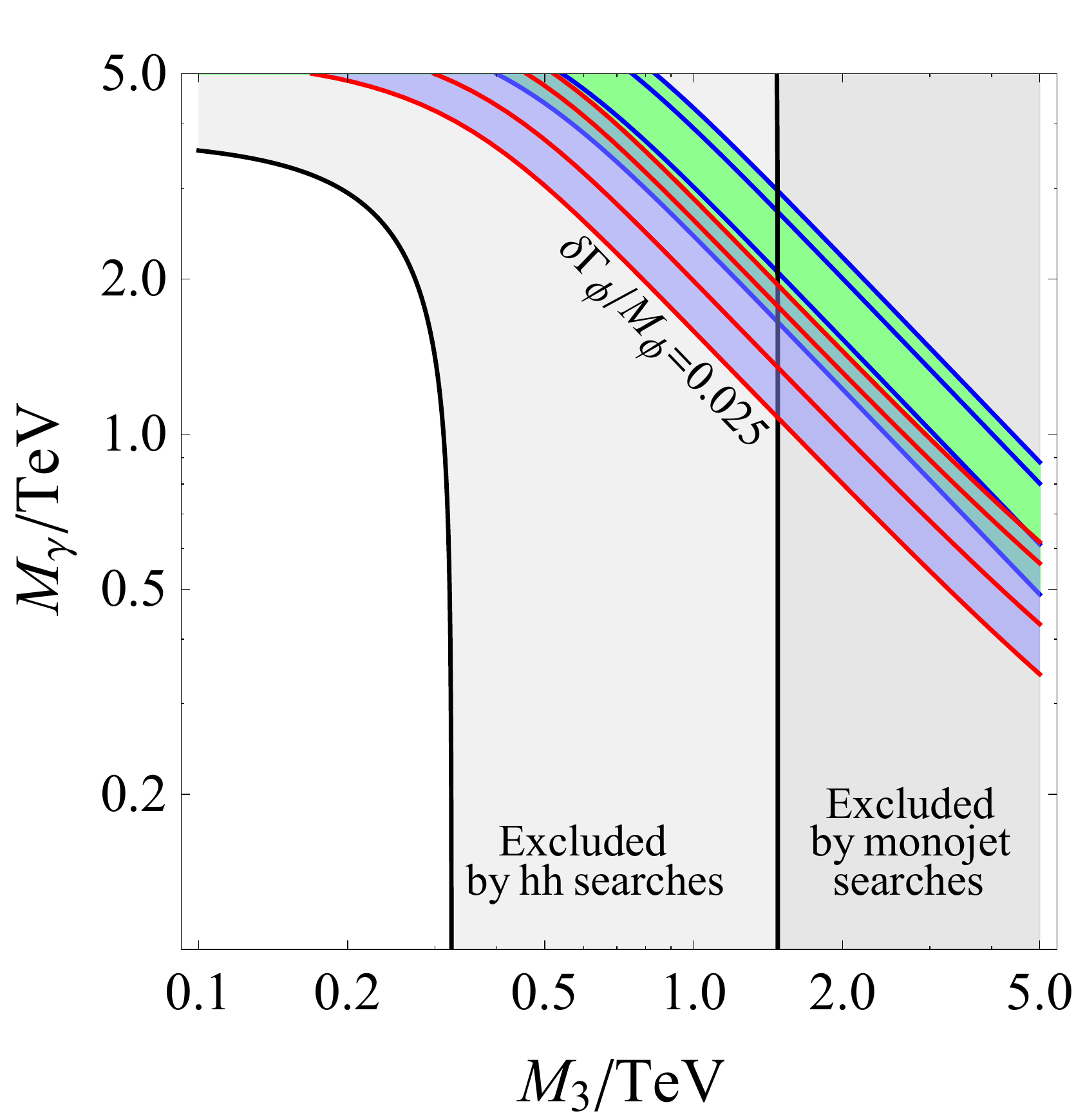} $$     
\begin{center}
\caption{Same as fig.~\ref{fig:M3Mph} but including an additional contribution to the sgoldstino partial width, $\delta \Gamma_\phi/M_\phi=0.025$ (blue band). The light (dark) gray region is excluded by LHC $hh$ (monojet) searches (if the additional partial widths is due to $\phi\rightarrow hh$ (invisible) decays). (The limit from dijet searches is displaced to the right of the plot in comparison with Fig. \ref{fig:M3Mph}).
\label{fig:M3Mphhh}}
 \end{center}
\end{figure}

On the other hand, the mixing between the original Higgs and the sgoldstino enables a new contribution to the $\phi\rightarrow hh$ decay. Namely, a term in the superpotential
\be
\delta W = \frac{c}{4!}\frac{F}{M^3} \Phi^4\ ,
\ee
induces a term  in the scalar potential
\be
\delta V=\frac{c}{3!}\frac{\tilde m^2}{ M}\Phi^3 +{\rm h.c.}
\ee
This gives a new contribution to the $\phi h^2$ coupling involved in $\phi\rightarrow hh$ decay, of size 
$\sim c \tilde m ^2 c_\alpha s^2_\alpha/M$ once mixing angle effects are taken into account.
Altough in principle this is parametrically smaller than the initial coupling in Eq.~(\ref{trilinear}), there is no mixing-angle obstruction to how large this new trilinear can be, so it can be substantially larger than $m_X$. Consequently the effective $\phi hh$ coupling, and thus the total width into Higgses, can be notably larger, eventually as large 
as suggested by ATLAS. However, having a large $\Gamma_{hh}$ can be in conflict with LHC $hh$ searches and one should further impose the limit $\Gamma_{hh}\simlt 20 (\Gamma_{\gamma\gamma})_{\rm obs}$ \cite{Strumia,ATLAShh}. Fig.~\ref{fig:M3Mphhh} shows how this constraint can ruin this as a solution to the large width problem.  In this figure, the region excluded by $hh$ searches assumes  $\delta \Gamma_{\phi,hh}/M_\phi=0.025$.

\section{Larger width from sgoldstino-Higgs mixing?\label{sec:shmix}}

As we have seen in the two previous sections, a trilinear coupling between sgoldstinos and Higgs, as in Eq.~(\ref{trilinear}), has two consequences: a splitting between the scalar and pseudoscalar components of the sgoldstino and a decay of sgoldstino into two Higgses.
Both effects can contribute to the apparent width of the 750 GeV resonance, as favoured by ATLAS data.

Here we discuss an additional effect of that mixing that enhances the sgoldstino width, but is potentially dangerous. Since the physical sgoldstino has a $\sin\alpha$ component of Higgs, the former can decay through the usual decay channels of the Higgs boson, with a rate suppressed by $\sin^2\alpha$. Now, for a 750 GeV  SM Higgs the decay is dominated by $WW$, $ZZ$ and $tt$, with the following partial widths\cite{Dittmaier:2011ti}: 
\bea
\Gamma(H(750\ {\rm GeV})\rightarrow WW) &=& 145\ {\rm GeV}\ ,
\nonumber\\
\Gamma(H(750\ {\rm GeV})\rightarrow ZZ)& = &71.9\ {\rm GeV}\ ,
\nonumber\\
\Gamma(H(750\ {\rm GeV})\rightarrow tt) &= &30.6\ {\rm GeV}\ .
\eea
Therefore the contribution to the total width of the sgoldstino is
\be
\delta \Gamma (\Phi\rightarrow WW, \ ZZ, \ tt)\simeq (247.5   \ {\rm GeV}) \sin^2\alpha
\label{extrawidth}
\ee
which is quite sizeable, even for mild mixing angles (it gives $\delta \Gamma \simlt 50$ GeV for $\sin^2\alpha\simlt 0.2$). In the last equation we have not considered the interference effects with the direct decays $\Phi\rightarrow WW,\ ZZ, \ tt$, which are typically subdominant.

There are two potentially dangerous drawbacks of this $\phi-h$ mixing. The first is that these enhaced $\Phi$ decays, particularly the one into $ZZ$, can be in conflict with LHC limits. Namely, one should respect the bound $\Gamma_{ZZ}\simlt 13 \Gamma_{\gamma\gamma}$  \cite{Ellis}. Using Eq.~(\ref{Gammas}) we get a bound on $\sin^2\alpha$:
\footnote{Given the interplay between the photino and gluino masses to accomodate the observed diphoton excess, this bound can be 
re-written as an upper bound on the gluino mass.}
\be
\sin^2\alpha \simlt 0.7\ \left(\frac{M_{\tilde\gamma}}{\sqrt{F}}\right)^2
\left(\frac{1\ {\rm TeV}}{\sqrt{F}}\right)^2\ .
\label{bound1}
\ee
Note that this bound restricts severely the possibility of a sgoldstino splitting due to mixing with the Higgs if the photino mass is substantially smaller that $\sqrt{F}$.

The second drawback is that the $\phi$ admixture in the Higgs will also affect the coupling of Higgs to gluons and photons (which are loop suppressed in the SM). Normalizing these couplings as $c_{gg}h/(4v)G_{\mu\nu}G^{\mu\nu}$ and $c_{\gamma\gamma}h/(4v)F_{\mu\nu}F^{\mu\nu}$, fits to LHC Higgs data put constraints  on $c_{\gamma \gamma}$  and $c_{gg}$ roughly of order $10^{-3}$, see e.g.~\cite{FRU}. The bound on $c_{\gamma \gamma}$  can be  used to set the constraint 
\be
 \sin^2\alpha \simlt 8\times 10^{-6}\left(\frac{\sqrt{F}}{M_{\tilde\gamma}}\right)^2
\left(\frac{\sqrt{F}}{1\ {\rm TeV}}\right)^2\ .
\label{bound2}
\ee
A similar bound on $\sin^2\alpha$, involving $M_3$ instead of $M_{\tilde\gamma}$, follows from the bound on $c_{gg}$.
Putting together Eq.~(\ref{bound1}) and (\ref{bound2}) sets an upper limit $\sin^2\alpha\simlt 2\times 10^{-3}$, and using this in Eq.~(\ref{extrawidth}) gives $\delta \Gamma_\phi\simlt 0.5$ GeV, a tiny shift, so this $h-\phi$ mixing mechanism cannot explain the large sgoldstino width.\footnote{The bound on $c_{\gamma\gamma}$ might be substantially weaker if $c_{\gamma \gamma}\simeq -2c_{\gamma \gamma}^{SM} $ (admittedly this would be a big coincidence). However, this requires  $M_{\tilde\gamma}/\sqrt{F}\simlt 1$, casting doubts on the EFT expansion.}

The arguments used in this section are of more general applicability and can constrain scenarios that mix the light Higgs and the scalar at 750 GeV (for work in this direction see \cite{Adam}).

\section{Sgoldstino decay into Higgsinos \label{sec:other}}

Besides the sgoldstino decay in two Higgs bosons, discussed in the previous section, the decay in two Higgsinos is an additional channel that could be naturally open and can be important.

From the superpotential in Eq.~(\ref{W})
one gets the interaction term
\be
\delta {\cal L} =  \frac{\mu'}{M}\ \Phi \tilde H_u\tilde H_d + \ {\rm h.c.}\ ,
\ee
between the sgoldstino and the Higgsinos, 
which allows $\Phi\rightarrow \tilde H \tilde H$ decay if $m_{\tilde H}\simeq\mu\leq M_\phi/2$. Provided the Higgsino is the LSP, this decay contributes to the invisible width of the sgoldstino.
LHC monojet searches constrain also such invisible decays, with the limit translating into $\Gamma_{inv}\simlt 400 (\Gamma_{\gamma\gamma})_{\rm obs}$ \cite{Strumia,monoj}. The impact of this limit is shown in Fig.~\ref{fig:M3Mphhh}. The contribution to the width is
\be
\Gamma_{\tilde H \tilde H}=\frac{M_\phi}{4\pi} \frac{{\mu'}^2}{M^2}\left(1-\frac{4m_{\tilde H}^2}{M_\phi^2}\right)^{3/2}\ .
\ee
Parametrically, using $\mu'\simlt \tilde{m}$, this width is of order $\tilde{m}^3/M^2$ like those discussed in section~\ref{sec:excess} but, being independent of gaugino masses, there is more freedom to increase it. Getting $\Gamma_{\tilde H \tilde H}/M_\phi=0.06$ requires $(\mu'/M)^2\simeq 0.84$ for $m_{\tilde H}\simgt 100$ GeV
 (its lower limit from LEP), a value that is too large to justify the 
effective theory expansion in powers of $\tilde{m}/M$. However, if we choose instead $(\mu'/M)^2=0.5$ we get $\Gamma_{\tilde H \tilde H}\simeq 27$ GeV, for the same value of $m_{\tilde H}$; a large value close enough to the ATLAS indication. Moreover, this is just a partial contribution to the width that should be added to others that could potentially be large, like that from the  $hh$ decay studied in section~\ref{sec:hh}. In addition, using this particular channel to enhance the sgoldstino width we do not run into the problem of clashing with LHC searches, as was the case for $\Gamma_{hh}$. In fact, fig.~\ref{fig:M3Mphhh} holds also if $\delta \Gamma_\phi$ is due to Higgsino decays, but now the excluded gray area (due to $hh$ searches) would not apply, and this leaves a region  (overlap between blue and green bands) that can succesfully explain the diphoton rate and the large width.

Finally, let us remark that the same operator that is responsible for the above sgoldstino coupling to Higgsinos also gives a (positive) contribution to the light Higgs mass through a $\lambda^{(F)}$ quartic coupling as discussed in section~\ref{sec:intro}, with
\be
\delta m_h^2 = \frac{1}{2}\frac{{\mu'}^2}{M^2} v^2 \sin^2 2\beta\ .
\ee
For $(\mu'/M)^2\sim 0.5$  one gets $\delta m_h^2\sim m_h^2\sin^2 2\beta$.
This can be very useful to reproduce the observed Higgs mass with less finetuning, one of the crucial virtues of this type of scenario \cite{CEH}. 

\section{Conclusions}

We have re-examined the diphoton excess observed by ATLAS and CMS \cite{data, Aad:2015mna,CMSgg8} as a possible supersymmetric signal of low-scale SUSY breaking (LSSB)  scenarios \cite{BCEN,CEH,Seiberg,otherstudies}. These models contain an excellent candidate to fit the signal: an scalar field (the sgoldstino) coupled to gluons and photons in a direct way, so that an effective production via gluon fusion and the subsequent decay into photons is possible. The partial widths into gluons (photons) depends on the ratio of the gluino (photino) mass over $\sqrt{F}$, i.e. the scale of SUSY breaking.

The possibility of accommodating the diphoton excess as a signal of LSSB has been proposed in \cite{Javi,Torre,ovs}. However, although the observed cross section is not difficult to fit, the typical width of the sgoldstino is much smaller than the value suggested by the ATLAS results, $\Gamma_\phi/M_\phi\simeq 0.06$. The authors of  
Ref.~\cite{Torre} presented a simple alternative explanation to this puzzle:
an splitting between the scalar and pseudoscalar degrees of freedom of the sgoldstino, which would be induced by a trilinear coupling between the sgoldstino and two Higgs fields, something typical in LSSB scenarios.
In this paper, we present an alternative mechanism to generate the sgoldstino splitting, which is very efficient and is not restricted by the strong bounds coming from the sgoldstino-Higgs mixing.

In this paper we also have explored other possibilities to  enhance the sgoldstino width, namely the decay into two Higgses, two Higgsinos and the contribution from mixing between the sgoldstino and the Higgs boson. The decay into Higgses arises from the above-mentioned trilinear couplings. 
 The maximal value of this partial width is extremely constrained by   (sgoldstino-Higgs) mixing effects. Typically, it turns out to be too small, although it is enhanced by the presence of trilinear sgoldstino operators, that are normal in LSSB. The mixing has other side effects, in particular it enables the decay of the sgoldstino through its Higgs-component, which enhances notably the total width. However, one must be careful not to violate the present bounds on $\Gamma_{ZZ}$,  as well as on Higgs couplings, particularly those from $h\rightarrow \gamma\gamma$ data. The combination of these two types of constraints imposes severe bounds on the scenario and, in particular, on the value of the mixing angle.
 
 Finally, the sgoldstino decay into Higgsinos can be very efficient if the latter are light enough. In summary there are interesting and very effective mechanisms to enhance the sgoldstino width, which, besides, lead to relevant predictions for LHC.

We find tantalizing that this (hint of a) signal  could correspond to an sgoldstino, a particle that lies at the very heart of supersymmetry breaking, similar in a sense to the central role of the Higgs for electroweak symmetry breaking. If nature is kind to us, this could represent a huge step forward in our understanding  of the origin of electroweak symmetry breaking and the role that supersymmetry presumably plays in it.

\section{Acknowledgments}
\label{ackn}
J.R.E. thanks Joan Elias-Mir\'o and Mario Mart\'{\i}nez for interesting discussions and acknowledges support by the Spanish Ministry MINECO under grants FPA2014-55613-P, FPA2013-44773-P, by the Generalitat de Catalunya grant 2014-SGR-1450 and by the Severo Ochoa excellence program of MINECO (grant SEV-2012-0234). J.R.E. and J.A.C. thank Hanna Jarzabek and Resu del Pozo for support during a very unusual Christmas. J.A.C. and J.M.M. are partially supported by the MINECO, Spain, under contract FPA2013-44773-P, Consolider-Ingenio CPAN CSD2007-00042, as well as MultiDark CSD2009-00064. We also thank the Spanish MINECO Severo Ochoa excellence program under grant SEV-2012-0249.


\end{document}